\documentclass{article}
\usepackage{emulateapj,onecolfloat}

\def\aa{{A\&A}}
\def\aas{{A\&AS}}
\def\aj{{AJ}}

\def\apj{{ApJ}}
\def\apjs{{ApJS}}

\def\mnras{{MNRAS}}
\def\nat{{Nature}}

\def\fun#1#2{\lower3.6pt\vbox{\baselineskip0pt\lineskip.9pt
\ialign{$\mathsurround=0pt#1\hfil##\hfil$\crcr#2\crcr\sim\crcr}}}
\def\lap{\mathrel{\mathpalette\fun <}}
\def\gap{\mathrel{\mathpalette\fun >}}
\def\mh{M_\bullet}

\begin{document}
\submitted{To appear in ``Coevolution 
of Black Holes and Galaxies,'' ed. L. C. Ho (Cambridge Univ. Press)}

\twocolumn
[
\title{Single and Binary Black Holes and their Influence 
on Nuclear Structure}
\author{DAVID MERRITT\\Rutgers University, New Brunswick, NJ 08903}

\begin{abstract}
Massive central objects affect both the structure and evolution
of galactic nuclei.
Adiabatic growth of black holes generates power-law central
density profiles with slopes in the range 
$1.5\lap -d\log\rho/d\log r \lap 2.5$,
in good agreement with the profiles observed
in the nuclei of galaxies fainter than $M_V\approx -20$.
However the shallow nuclear profiles of bright galaxies
require a different explanation.
Binary black holes are an inevitable result of galactic mergers,
and the ejection of stars by a massive binary 
displaces a mass of order the binary's own mass,
creating a core or shallow power-law cusp.
This model is at least crudely consistent with core
sizes in bright galaxies.
Uncertainties remain about the effectiveness
of stellar- and gas-dynamical processes at inducing
coalescence of binary black holes, and uncoalesced binaries
may be common in low-density nuclei.
Numerical $N$-body experiments are not well suited to
probing the long-term evolution of black hole binaries 
due to spurious relaxation.
\end{abstract}
]

\section{Introduction}

The effect of a supermassive black hole on its stellar surroundings
depends, as so often in stellar dynamics, on how one imagines the
system evolved to its present state. Collisional relaxation times
are too long to have affected the stellar distribution in all but
the densest nuclei (Faber et al. 1997), hence the structure and 
kinematics of nuclei are fossil relics of the interactions between 
stars and black holes. In the simplest scenario, the black hole
grows by accreting gas on a time scale long compared
with the orbital periods of the surrounding stars
(Peebles 1972; Young 1980).
This ``adiabatic
growth'' model makes fairly definite predictions about the distribution
of stars near the black hole, predictions which are consistent
with the steep density cusps observed in faint galaxies but
which can not explain the flatter profiles at the
centers of bright galaxies.  But galaxies merge, implying
the formation of binary black holes
(Begelman, Blandford \& Rees 1980) which are efficient at
displacing matter as they spiral together.  The dynamics of black hole
binaries in galactic nuclei are complex; 
among the unanswered questions are the long-term
efficiency of stellar dynamical processes at extracting energy
from a binary, and whether decay of black hole binaries
ever stalls at separations too great for the efficient emission
of gravitational waves.
But the binary black hole model is at least crudely
consistent with the observed dependence of nuclear
structure on galaxy luminosity.
This article summarizes theoretical work on the single and
binary black hole models and suggests avenues for
future progress.

\section{Preliminaries}

A black hole of mass $M_\bullet$ embedded in a galactic nucleus will
strongly affect the motion of stars within a distance $r=r_h$,
the ``radius of influence.''
A standard definition for $r_h$ is
\begin{equation}
r_h\equiv{G\mh\over\sigma^2} \approx 10.8\ {\rm pc} \left({M_\bullet\over 10^8M_\odot}\right) \left({\sigma\over 200\ {\rm km\ s}^{-1}}\right)^{-2}
\label{eq:rh}
\end{equation}
with $\sigma$ the 1D velocity dispersion of the stars
at $r\gg r_h$.
This definition had its origin in the isothermal sphere model
for galactic nuclei; $\sigma$ is independent of $r$ in
such a model and $r_h$ is the radius
at which the circular velocity around the black hole equals
$\sigma$.
We now know that nuclei are power laws in the
stellar density, $\rho\sim r^{-\gamma}$, and that
$\gamma$ can lie anywhere between $\sim 0$ and $\sim 2.5$
(Lauer, this volume).
The velocity dispersion in a power-law nucleus
is only constant if $\gamma=0$ or $2$,
hence a definition like equation (\ref{eq:rh}) is problematic.
One alternative would be to define $r_h$ as the root of
$\sigma^2(r)-G\mh/r=0$.
A simpler definition, which will be adopted in this article,
is the radius at which the enclosed mass
in stars is twice the black hole mass:
\begin{equation}
M_*(r<r_h) = 2\mh.
\label{eq:def_rh}
\end{equation}
This definition is exactly equivalent to equation (\ref{eq:rh}) 
when $\rho(r)=\sigma^2/2\pi Gr^2$,
the singular isothermal sphere.
For an arbitrary power law,
\begin{equation}
\rho(r) = \rho_0\left({r\over r_0}\right)^{-\gamma},
\label{eq:power}
\end{equation}
equation (\ref{eq:def_rh}) implies
\begin{equation}
r_h = r_0\left({3-\gamma\over 2\pi}{\mh\over\rho_0r_0^3}\right)^{1/(3-\gamma)}.
\end{equation}
Note that the ``theorist's'' convention is adopted here, 
in which $\gamma$ is the power-law index of the space
(not projected) density.
 
\section{The Adiabatic Growth Model}

If a black hole grows at the center of a stellar system 
through the accretion of gas, the stellar density in the core
will also grow as the black hole's gravity pulls in 
nearby stars (Peebles 1972; Young 1980).
The change in the stellar density can be computed
straightforwardly if it is assumed that the black
hole grows on a time scale long compared with
stellar orbital periods.
This is reasonable, since even Eddington-limited accretion
requires $\sim 10^8$ yr to double the black hole mass,
and orbital periods throughout the region dominated by
the black hole are $\lap 10^6$ yr.
Under these assumptions, the adiabatic invariants ${\mathbf J}$ 
associated with the stellar orbits are conserved as the black hole grows
and the phase-space density $f$ remains fixed when
expressed in terms of the ${\mathbf J}$.
Computing the final $f$ becomes a simple matter of
expressing the final orbital integrals in terms of their
initial values under the constraint that the adiabatic
invariants remain fixed (Young 1980).

In spherical potentials, the adiabatic invariants 
are the angular momentum $L$ and the radial action 
$I=2\int_{r_-}^{r_+}\sqrt{2\left[E-\Phi(r)\right]-L^2/r^2}$,
where $\Phi(r)$ is the gravitational potential and $r_\pm$ are 
pericenter and apocenter radii.
It may be shown (e.g. Lynden-Bell 1963)
that orbital shapes remain nearly unchanged when $L$ and $I$ 
are conserved, implying that 
an initially isotropic velocity distribution $f(E)$ 
remains nearly isotropic after the black hole grows
(though not exactly isotropic -- see below).
The final $f$ corresponding to an initially isotropic $f$ is then simply
\begin{eqnarray}
f_f(E_f,L)&=&f_i(E_i,L) \nonumber \\
&\approx& f_f(E_f)
\end{eqnarray}
where $E_f$ is related to $E_i$ through the condition
$I_f(E_f,L)=I_i(E_i,L)$.

\begin{figure*}[t]
\includegraphics[width=3.5in,angle=-90.]{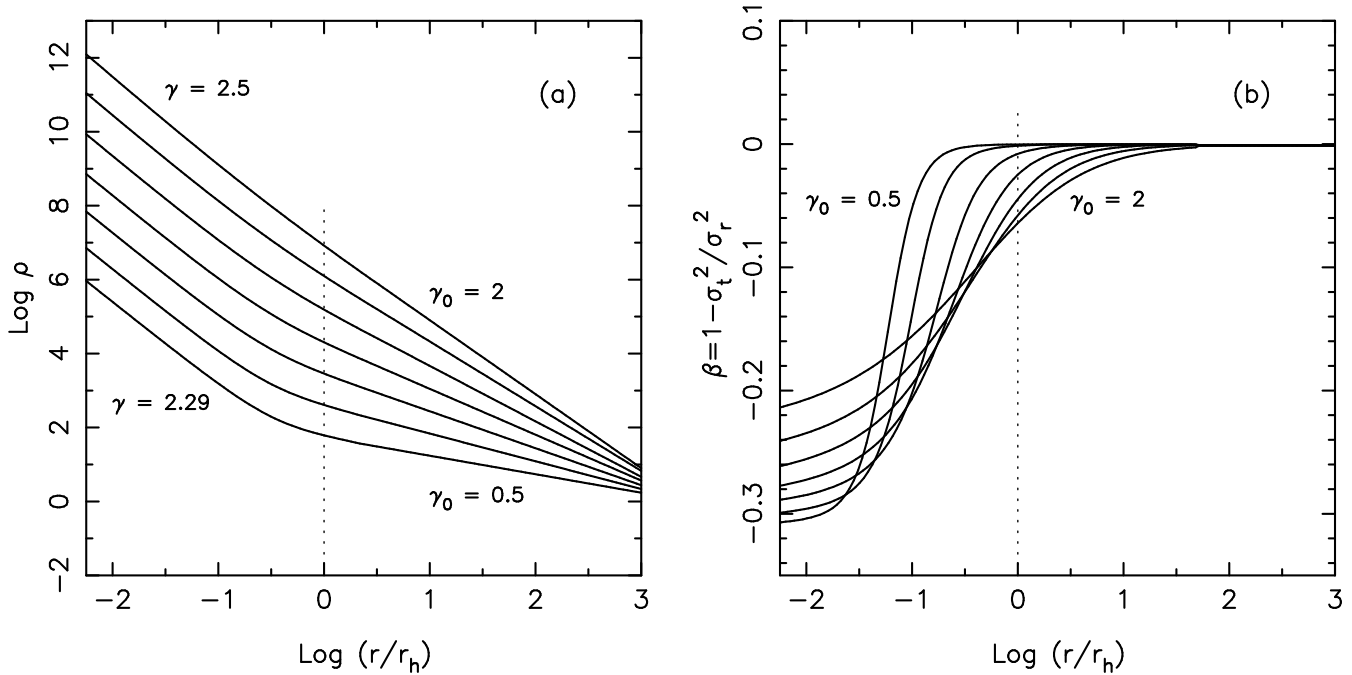}
\epsscale{1.0}
\caption{
Influence of the adiabatic growth of a black hole on
its nuclear environment in a spherical, isotropic galaxy.
(a) Density profiles after growth of the black hole.
Initial profiles were power laws, $\rho_i\propto r^{-\gamma_0}$,
with $\gamma_0$ increasing upwards in steps of $0.25$.
The radial scale is normalized to $r_h$ as defined in the initial
galaxy (eq. \ref{eq:def_rh}).
The slope of the final profile at $r< r_h$ is 
almost independent of the initial slope.
(b) Velocity anisotropies after growth of the black hole.
A slight bias toward circular motions appears at $r< r_h$.}
\label{fig:one}
\end{figure*}

While the form of the nuclear density profile before the
black hole appeared is not known,
$N$-body studies of structure formation
suggest that power laws are generic 
(Power et al. 2003 and references therein)
and observed nuclear density profiles are often well described
as power laws even on scales $r\gg r_h$.
Setting $\rho_i\propto r^{-\gamma_0}$, 
$\Phi_i\propto r^{2-\gamma_0}$ ($0<\gamma_0<2$),
the initial distribution function becomes
\begin{equation}
f_i(E_i) \propto E_i^{-\beta},\ \ \ \ \beta={6-\gamma_0\over 2(2-\gamma_0)} \ \ \ \ (0<\gamma_0<2).
\end{equation}
To compute $f_f(E_f)$ we need a relation between $E_f$ and
$E_i$; we restrict attention to the region within the black hole's
sphere of influence by setting 
$E_f=v^2/2 -GM_\bullet/r$.
The radial action $I(E,L)$ in the power-law model can not
be computed analytically for every $(E,L)$,
but for certain orbits $E_f(E_i)$ has a simple form.
For instance, circular orbits have $I=0$, and conservation
of angular momentum implies $r_iM_i(r_i) = r_fM_\bullet$
or $r_f\propto r_i^{4-\gamma_0}$.
Thus $E_f\propto -r_f^{-1}\propto -r_i^{\gamma_0-4}\propto 
-E_i^{(\gamma_0-4)/(2-\gamma_0)}$,
or
\begin{equation}
E_i\propto (-E_f)^{-(2-\gamma_0)/(4-\gamma_0)}.
\end{equation}
The same relation turns out to be precisely correct for radial orbits
as well and is nearly correct at intermediate eccentricities
(Gondolo \& Silk 1999).
Thus we can write
\begin{equation}
f_f(E_f) =f_i(E_i) \propto E_i^{-\beta}\propto (-E_f)^\delta,\ \ \ \ \delta={6-\gamma_0\over 2(4-\gamma_0)}
\end{equation}
and the final density profile within the sphere of influence of the black
hole is
\begin{eqnarray}
\rho_f(r) &=& \int f_f(v)d^3 v
\propto \int_{\Phi(r)}^0 (-E)^\delta\sqrt{E-\Phi(r)}\ dE \nonumber \\
&\propto& r^{-\gamma},\ \ \ \ \gamma = 2+{1\over 4-\gamma_0}.
\end{eqnarray}
For $0<\gamma_0<2$, $\gamma$ varies only between $2.25$ and $2.5$;
the slope of the final density profile within $r_h$ 
is almost independent of $\gamma_0$.

The form of $\rho_f(r)$ at $r\approx r_h$
must be computed numerically (e. g. Young 1980; Cipollina \& Bertin 1994;
Cipollina 1995; Quinlan, Hernquist \& Sigurdsson 1995; Gondolo \& Silk 1999).
Figure \ref{fig:one} shows $\rho_f(r)$ when $\rho_i(r)\propto r^{-\gamma_0}$.
Defining $r_{cusp}$ to be the radius at which the inner and outer
power laws intersect, one finds
\begin{equation}
r_{cusp} = \alpha r_h,\ \ \ \ 0.19\lap\alpha\lap 0.22,\ \ \ \ 0.5\le\gamma_0\le 1.5.
\end{equation}

In early treatments of the adiabatic growth model
(Peebles 1972; Young 1980),
the black hole was assumed to grow inside of a constant-density
isothermal core.
The index of the power-law cusp that forms from this initial state
is $\gamma=1.5$, compared with the limiting valure $\gamma=2.25$ 
as $\gamma_0\rightarrow 0$ in the power-law models.
This difference can be traced to differences in the 
central density profile:
\begin{equation}
\rho_i(r) = \rho_0\times\left(1 + C_1r + C_2 r^2 + ...\right).
\end{equation}
The isothermal model has $C_1=0$ (an ``analytic core'')
implying a phase space density that tends to a constant value
at low energies.
Other sorts of cores have $C_1\ne 0$ and 
$f$ diverges at low energies; for instance,
the core produced by setting $\gamma=0$ in 
$\rho(r) = r^{-\gamma}(1+r)^{-4}$ has
$f(E)\rightarrow [E-\Phi(0)]^{-1}$.
In fact models with finite central $\rho$'s can be found
that generate final cusp slopes anywhere in the
range $1.5\le\gamma\le 2.25$ (Quinlan, Hernquist \& Sigurdsson 1995).
There is probably no way of ruling out an analytic core 
in the progenitor galaxy on the very small scales that are
relevant to the later formation of a cusp,
hence the adiabatic growth model is compatible with
any final slope in the range $1.5\lap\gamma\lap 2.5$.
The upper limit could even be extended beyond $2.5$ if
$\gamma_0>2$.

How do these predictions compare with the data?
Observed luminosity profiles are well described 
as power laws at the smallest resolvable radii,
and in the case of faint ellipticals,
$M_V\gap -20$, the observed range of slopes 
is $1.5\lap\gamma\lap 2.5$ (Lauer, this volume).
This is precisely the range in $\gamma$ predicted by the
adiabatic growth model.
However in bright galaxies, $\gamma$ extends down to 
$\sim 0$.
A natural interpretation is that the steep cusps 
in faint galaxies are a result of adiabatic
black hole growth, while some additional mechanism,
like mergers,
has acted to modify the profiles in the brighter galaxies.

Some fine tuning is still required to reproduce
the luminosity profiles of galaxies with $1.5\lap\gamma\lap 2$.
These intermediate slopes require a shallow, core-like initial profile,
and if the initial core radius exceeded $\sim r_h$,
the final profile will exhibit an upward inflection
at $r\approx r_h$ (e.g. Figure 2 of Young 1980; Figure \ref{fig:one}).
Such inflections are rarely if ever seen; 
observed profiles have slopes that decrease smoothly inward.
A way out is to require that the black hole
mass exceeds the initial core mass so that its growth
obliterates the core; alternatively,
all nuclei with $\gamma\lap 2$ may have been the products
of mergers.

An ingenious attempt to reconcile the adiabatic growth
model with both steep and shallow cusps 
was made by van der Marel (1999).
He postulated the existence of isothermal cores in the
progenitor galaxies with core masses scaling as 
$\sim L^{1.5}$, with $L$ the total galaxy luminosity.
Since $\mh\sim L$, $\mh/M_{core}\sim L^{-0.5}$ and
the black holes in faint galaxies would grow to
dominate their cores, producing a cusp profile that
approximates the featureless power laws of faint galaxies.
In bright galaxies, van der Marel argued that the upward
inflection at $r\lap r_h$ would be difficult to resolve;
hence these galaxies would exhibit nearly unperturbed core
profiles.
Van der Marel's model is intriguing, although the assumed
relation between core size and galaxy luminosity is ad hoc,
and the assumption of
large, pre-existing cores does not fit naturally into
any current model of galaxy formation.

The predictions of the adiabatic growth model can be altered
if the initial galaxy is anistropic, non-spherical or rotating.
The simplest such case to treat is a spherical nucleus
containing only circular orbits, with or without rotation
(Young 1980; Quinlan, Hernquist \& Sigurdsson 1995;
Ullio, Zhao \& Kamionkowski 2001).
The derivation given above applies without approximations to this case:
if the initial density profile is a power law, 
$\rho_i\propto r^{-\gamma_0}$,
the final profile 
is also a power law with $\gamma=2+1/(4-\gamma_0)$,
for any $\gamma_0$ in the range $[0,3]$.
As shown above, the isotropic model yields approximately the same final slope
for $0<\gamma_0<2$,
suggesting that even extreme tangential anisotropies have little
effect on the final profile.
The effect of radial anistropies has apparently never been 
investigated; this is worth doing since 
real galaxies often show evidence for significant 
radial anisotropies (e.g. Kronawitter et al. 2000).

Black hole growth in axisymmetric nuclei with analytic cores
was considered by Leeuwin \& Athanassoula (2000).
They found little dependence of the final cusp slope on
the degree of flattening.
Merritt \& Quinlan (1998) grew black holes of various masses
in a triaxial $N$-body model that was formed via gravitational
collapse, and found $\rho\sim r^{-2}$ at $r\lap r_h$.
Their models evolved to axisymmetry after growth of the black
hole, but it is now known that triaxiality can be maintained
throughout the black hole's sphere of influence
(Poon \& Merritt 2002, 2003).
Triaxial potentials support a wide range of different orbit 
families and the effects of black hole growth on such models
have not been examined in detail.

In the spherical geometry, adiabatic growth of a black hole induces
a mild anisotropy in the stellar motions at $r\lap r_h$ due to the
slightly different ways that circular and eccentric orbits respond to the
changing potential; the net effect is a decrease in the
average orbital eccentricity
(Young 1980; Goodman \& Binney 1984; Quinlan, Hernquist
\& Sigurdsson 1995; Figure~\ref{fig:one}).
If the progenitor galaxy is rotating,
growth of the black hole tends to increase $V_{rot}$ more
rapidly than $\sigma$ (Lee \& Goodman 1989; Leeuwin \& Athanassoula
2000), although again the effect is slight.

To summarize: the adiabatic growth model is limited in its ability
to reproduce the full range of luminosity profiles
observed in galactic nuclei.
The model predicts power-law profiles at $r\lap r_h$ with
logarithmic slopes $1.5\lap\gamma\lap 2.5$.
This nicely brackets the range of slopes observed in the nuclei of galaxies
fainter than $M_V\approx -20$.
However slopes less than $\gamma\approx 1.5$ are not naturally produced
by the adiabatic growth model, and some fine tuning is required
to avoid inflections in the profile at $r\approx r_h$.

\begin{figure*}[t]
\includegraphics[width=4.in,angle=-90.]{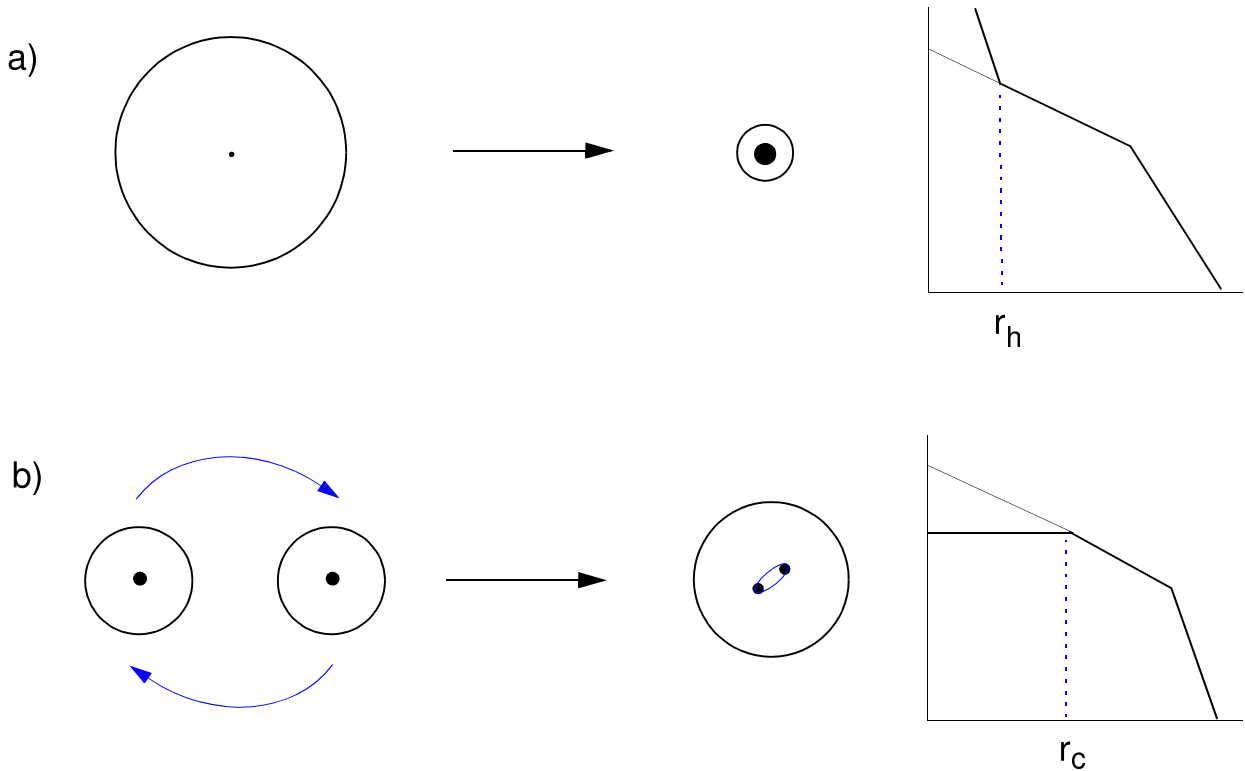}
\caption{
Two schemes for growing black holes at the centers of galaxies.
(a) The adiabatic growth model; the stellar density within
$r_h$ is increased as the black hole pulls in stars. 
(b) The binary black hole model; the inspiralling black holes
displace matter within a distance $r_c$ that is roughly
the separation between the two black holes when they first
form a bound pair.}
\label{fig:two}
\end{figure*}

\section{The Binary Black Hole Model}

The adiabatic growth model was proposed (Peebles 1972)
before the importance of galaxy interactions and mergers 
was appreciated.
We now know that supermassive black holes have been present in at least some
spheroids since redshifts of $z\approx 6$ (e.g. Fan et al. 2001),
and we believe that most galaxies have experienced at least
one major merger since that time; indeed the era of
peak quasar activity may coincide with the era of galaxy
assembly via mergers (e.g. Caviliere \& Vittorini 2000).
If a nucleus forms via the merger of two galaxies
containing pre-existing black holes, the net effect on the nuclear density 
profile is roughly the opposite of what 
the adiabatic growth model predicts: the black holes {\it displace}
matter as they spiral into the center (Figure \ref{fig:two}).
This is a natural way to account for the shallow nuclear
profiles in bright galaxies.
The process may be understood as a sort of dynamical friction,
with the ``heavy particles'' (the black holes) transferring their
kinetic energy to the ``light particles'' (the stars).
However most of the energy transfer
takes place after the two black holes have come within each
other's sphere of influence, and in this regime, the interaction
with the background is dominated by another mechanism,
the {\it gravitational slingshot} (Saslaw, Valtonen \& Aarseth 1974).
The massive binary ejects passing stars at high velocities,
removing them from the nucleus and simultaneously
increasing its binding energy.

We begin by reviewing the dynamics of massive binaries in
fixed, homogeneous backgrounds, then consider the more difficult 
problem of a binary located at the center of an inhomogeneous 
and evolving galaxy.

\subsection{Dynamics of Massive Binaries}
\label{sec:dmbs}

Consider a binary system consisting of two black holes
with masses $M_1$ and $M_2$, where $p\equiv M_2/M_1\le 1$ and 
$M_{12}\equiv M_1+M_2$.
Let $a$ be the semi-major axis of the Keplerian orbit
and $e$ the orbital eccentricity.
The binding energy of the binary is
\begin{equation}
|E| = {GM_1M_2\over 2a} = {G\mu M_{12}\over 2a},
\end{equation}
with $\mu=M_1M_2/M_{12}$ the reduced mass.
The relative velocity of the two black holes,
assuming a circular orbit, is
\begin{equation}
V_{bin} = \sqrt{GM_{12}\over a} = 658\ {\rm km\ s}^{-1} \left({M_{12}\over 10^8 M_\odot}\right)^{1/2} \left({a\over 1\ {\rm pc}}\right)^{-1/2}.
\end{equation}
A binary is called ``hard'' when its binding energy per unit
mass, $|E|/M_{12} = G\mu/2a$, exceeds $\sim\sigma^2$.
For concreteness, a binary will here be called hard if
its separation falls below $a_h$, where
\begin{equation}
a_h \equiv {G\mu\over 4\sigma^2} \approx 0.27\ {\rm pc} \left(1+p\right)^{-1} \left({M_2\over 10^7 M_\odot}\right) \left({\sigma\over 200\ {\rm km\ s}^{-1}}\right)^{-2}.
\end{equation}
Other definitions of $a_h$ are possible (e. g. Hills 1983; Quinlan 1996).

Stars passing within a distance $\sim 3a$ of the center of mass
of a hard binary undergo a complex interaction with the two
black holes, followed almost always by ejection at velocity
$\sim \sqrt{\mu/M_{12}} V_{bin}$
(Saslaw, Valtonen \& Aarseth 1974).
Each ejected star carries away energy and angular momentum,
causing the semi-major axis, eccentricity and orientation of the
binary to change and the local density of stars to drop.
If the stellar distribution is assumed fixed far from the binary
and if the contribution to the potential from the stars is ignored, 
the rate at which these changes occur
can be computed by carrying out scattering experiments
of massless stars against a binary whose 
orbital elements remains fixed during each interaction.

The results of the scattering experiments can be summarized
via a set of dimensionless coefficients $H, J, K, L, ...$
which define the mean rates of change of the parameters
characterizing the binary and the stellar background
(Hills \& Fullerton 1980; Roos 1981; Hills 1983, 1992;
Baranov 1984; Mikkola \& Valtonen 1992; Quinlan 1996; Merritt 2001, 2002).
These coefficients are functions of the binary mass ratio,
eccentricity and hardness but are typically independent
of $a$ in the limit that the binary is very hard.
The hardening rate of the binary is given by
\begin{equation}
{d\over dt}\left({1\over a}\right) = H{G\rho\over\sigma}
\label{eq:def_H}
\end{equation}
with $\rho$ the density of stars at infinity.
The mass ejection rate is
\begin{equation}
{dM_{ej}\over d\ln(1/a)} = J M_{12}
\label{eq:def_J}
\end{equation}
with $M_{ej}$ the mass in stars that escape the binary.
The rate of change of the binary's orbital eccentricity is
\begin{equation}
{de\over d\ln(1/a)} = K.
\label{eq:def_K}
\end{equation}
The diffusion coefficient describing changes in the
binary's orientation is
\begin{equation}
{\langle\Delta\vartheta^2\rangle}= L{m_*\over M_{12}}{G\rho a\over\sigma}
\label{eq:def_L}
\end{equation}
with $m_*$ the stellar mass.
Additional coefficients describe the rate of diffusion of the
binary's center of mass, or ``Brownian motion'' (Merritt 2001).

The binary hardening coefficient $H$ reaches a constant value
of $\sim 16$ in the limit $a\ll a_h$, with a weak dependence
on $M_2/M_1$
(Hills 1983; Mikkola \& Valtonen 1992; Quinlan 1996).
In a fixed background, equation (\ref{eq:def_H}) therefore implies that
a hard binary hardens at a constant
rate:
\begin{equation}
{1\over a(t)} - {1\over a_h} \approx 
H {G\rho \over \sigma}\left(t-t_h\right),\ \ \ \ t\ge t_h, \ \ \ \ a(t_h)=a_h.
\label{eq:decay}
\end{equation}
If the supply of stars remains steady,
hardening continues at this rate until the components of 
the binary come close enough together that the emission
of gravitational radiation is important.
In this regime, gravity wave coalescence takes place in a time:
\begin{eqnarray}
t_{gr} &=& {5\over 256F(e)} {c^5\over G^3\mu M_{12}^2}a^4, \nonumber \\
F(e) &=& (1-e^2)^{-7/2}\left(1 + {73\over 24}e^2 + {37\over 96}e^4\right)
\label{eq:t_gr}
\end{eqnarray}
(Peters 1964).
Coalescence in a time $t_{gr}$ occurs when $a=a_{gr}$, where
\begin{equation}
{a_h\over a_{gr}} \approx 75 F^{-1/4}{p^{3/4}\over\left(1+p\right)^{3/2}} 
\left({\sigma\over 200\ {\rm km\ s}^{-1}}\right)^{-7/8} 
\left({t_{gr}\over 10^9{\rm yr}}\right)^{-1/4}.
\label{eq:a_gr}
\end{equation}
The $M_\bullet-\sigma$ relation has been used to 
express $M_{12}$ in terms of $\sigma$.
For mass ratios $p$ of order unity and $e\approx 0$, 
the binary must 
decay by a factor of $\sim 10^2$ in order
for gravitational radiation to induce coalescence
in a time shorter than $10^9$ yr.
Less decay is required if the binary is eccentric
or if $M_2\ll M_1$.

\begin{figure}[t]
\includegraphics[width=3.5in,angle=0.]{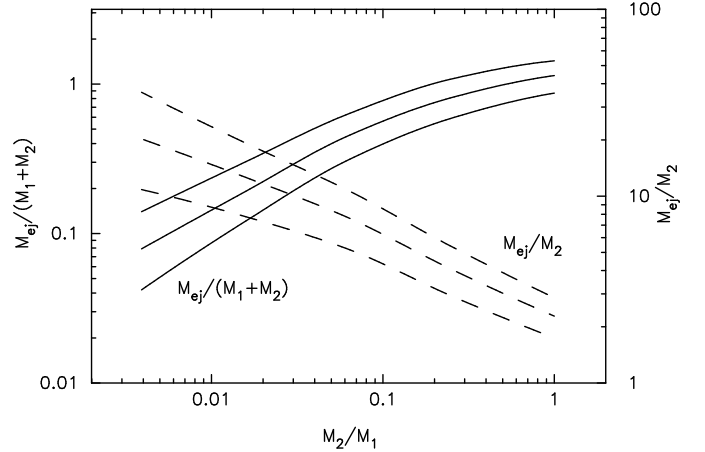}
\caption{
Mass ejected by a decaying binary, in units of 
$M_{12}=M_1+M_2$ (solid lines) or $M_2$ (dashed lines).
Curves show mass that must be ejected in order for the binary
to reach a separation where the emission of gravity
waves causes coalescence on a time scale of 
$10^{10}$ yr (lower), $10^9$ yr (middle) and $10^8$ yr (upper).}
\label{fig:three}
\end{figure}

If the binary manages to shrink by such a large factor,
the damage done to its stellar surroundings will be considerable.
The mass ejected by the binary in decaying from $a_h$ to $a_{gr}$
is given by the integral of equation (\ref{eq:def_J}):
\begin{equation}
M_{ej} = M_{12}\int_{a_{gr}}^{a_h} {J(a)\over a} da.
\label{eq:mej_int}
\end{equation}
Figure 1.2 shows $M_{ej}$ as a function of the mass ratio 
$M_2/M_1$ for $\sigma=200$ km s$^{-1}$ and various values
of $t_{gr}$.
The mass ejected in reaching coalescence is of order
$M_{12}$ for equal-mass binaries, and several times $M_2$
when $M_2\ll M_1$.
A black hole that grew to its current size through
a succession of mergers should therefore have displaced
a few times its own mass in stars.
If this mass came mostly from stars that were originally 
in the nucleus,
the density within $r_h$ would drop drastically and the
hardening would slow.
Without some way of replenishing the supply of stars,
decay could stall at a separation much greater
than $a_{gr}$.
This is the ``final parsec problem'': 
how to avoid stalling and bring the 
black holes from their separation of $\sim 1$ pc,
when they first form a hard binary, 
to $\sim 10^{-2}$ pc, where gravity wave emission
is efficient.

Changes in the binary's orbital eccentricity
(equation \ref{eq:def_K})
are potentially important because the gravity wave coalescence
time drops rapidly as $e\rightarrow 1$ (equation \ref{eq:t_gr}).
For a hard binary, scattering experiments give
$K(e)\approx K_0 e (1-e^2)$, with $K_0\approx 0.5$ for an
equal-mass binary
(Mikkola \& Valtonen 1992; Quinlan 1996).
The dependence of $K$ on $M_2/M_1$ is not well understood.
The implied changes in $e$ as a binary decays from $a=a_h$ to $a_{gr}$
are modest, $\Delta e\lap 0.2$, for all initial eccentricities.
The rms change in the orientation of the binary's spin axis
(equation \ref{eq:def_L}) is 
$\delta\theta\approx 2(m_*/M_{12})^{1/2}\log^{1/2}(a_h/a)$,
which is of order one degree or less if $m_*\approx M_\odot$.
Reorientations of the binary's angular momentum affect the spin
direction of the coalesced black hole and the direction
of any associated jet (Merritt 2002).

\subsection{Binary Black Holes in Galaxies}
\label{sec:bbh_ingals}

The scattering experiments summarized above treat the
binary's environment as fixed and homogeneous.
In reality, the binary is embedded at the center of
an inhomogeneous and evolving galaxy, and the supply of stars 
that can interact with it is limited.

In a fixed spherical galaxy,
stars can interact with the binary only if their pericenters
lie within $\sim {\cal R}\times a$, where ${\cal R}$ is of order
unity.
Let $L_{lc}={\cal R}a\sqrt{2\left[E-\Phi({\cal R}a)\right]}\approx \sqrt{2GM_{12}{\cal R}a}$,
the angular momentum of a star with pericenter ${\cal R}a$.
The ``loss cone'' is the region in phase space defined by
$L\le L_{lc}$.
The mass of stars in the loss cone is
\begin{eqnarray}
M_{lc}(a)&=&m_*\int dE \int_0^{L_{lc}} dL\ N(E,L^2) \nonumber \\
&=& m_*\int dE\int_0^{L_{lc}^2} dL^2 4\pi^2 f(E,L^2) P(E,L^2) \nonumber \\
&\approx& 8\pi^2 GM_{12}m_* {\cal R}a\int dE f(E) P_{rad}(E).
\label{eq:def_mlc}
\end{eqnarray}
Here $P$ is the orbital period; in the final line,
$f$ is assumed isotropic and $P$ has been
approximated by the period of a radial orbit of energy $E$.
An upper limit to the mass that is available to interact with 
the binary 
is $\sim M_{lc}(a_h)$, the mass within the loss cone when the
binary first becomes hard; 
this is an upper limit since some stars that are initially within
the loss cone will ``fall out'' as the binary shrinks.
Assuming a singular isothermal sphere for the stellar distribution,
$\rho\propto r^{-2}$, and taking the lower limit of the energy
integral to be $\Phi(a_h)$, 
equation (\ref{eq:def_mlc}) implies
\begin{equation}
M_{lc}(a_h) \approx 3{\cal R}\mu.
\label{eq:mlc}
\end{equation}
We can compute the change in $a$ that would result if
the binary interacted with this entire mass, by using
the fact the mean energy change of a star interacting 
with a hard binary is $\sim 3G\mu/2a$ (Quinlan 1996).
Equating the energy carried away by stars with
the change in the binary's binding energy gives
\begin{equation}
{3\over 2}{G\mu\over a} dM \approx {GM_1M_2\over 2}d\left({1\over a}\right)
\label{eq:dm1}
\end{equation}
or
\begin{equation}
\ln\left({a_h\over a}\right) \approx 3{\Delta M\over M_{12}} \approx
{9{\cal R}\mu\over M_{12}}\approx 9{\cal R} {p\over (1+p)^2}, \ \ \ \ \ \ p\equiv M_2/M_1
\label{eq:dm2}
\end{equation}
if $\Delta M$ is equated with $M_{lc}$.
Only for very low mass ratios ($M_2\lap 10^{-3}M_1$) 
is this decay factor large enough to satisfy
 equation (\ref{eq:a_gr}), but the time required for 
such a small black hole
to reach the nucleus is likely to exceed a Hubble time
(Merritt 2000).
Hence even under the most favorable assumptions, the
binary would not be able to interact with enough mass
to reach gravity-wave coalescence.

But the situation is even worse than this, since
not all of the mass in the loss cone will find its
way into the binary.
The time scale for the binary to shrink is comparable with
stellar orbital periods, and some of the stars with $r_{peri}\approx a_h$
will only reach the binary after $a$ has fallen below $\sim a_h$.
We can account for the changing size of the loss cone by writing
\begin{eqnarray}
{dM\over dt} &=& \int_{E_0(t)}^{\infty} {1\over P(E)} {d M_{lc}\over dE} dE 
\nonumber \\ 
&=& 8\pi^2 GM_{12}m_* {\cal R}a(t)\int_{E_0(t)}^{\infty} f_i(E)dE,
\label{eq:dm3}
\end{eqnarray}
where $M(t)$ is the mass in stars interacting with the binary and
$f_i(E)$ is the initial distribution function; setting
$P(E_0)=t$ reflects the fact that stars on
orbits with periods less than $t$ have already interacted with the binary
and been ejected.
Combining equations (\ref{eq:dm1}) and (\ref{eq:dm3}),
\begin{equation}
{d\over dt}\left({1\over a}\right) \approx 24\pi^2{\cal R}G m_* \int_{E_0(t)}^{\infty} f_i(E) dE.
\label{eq:shrink}
\end{equation}
Solutions to equation (\ref{eq:shrink}) show that a binary in
a singular isothermal sphere galaxy stalls at $a_h/a\approx 2.5$ for
$M_2= M_1$, compared with $a_h/a\approx 10$ if the full loss cone
were depleted (equation~\ref{eq:dm2}).
In galaxies with shallower central cusps, decay of the binary would stall
at even greater separations.

It is therefore entirely possible that uncoalesced binaries exist 
at the centers of many galaxies, particularly galaxies like large ellipticals
with low central densities.
But there is circumstantial evidence that long-lived black hole binaries
are rare.
Some radio galaxies show clear evidence of a recent flip in
the black hole's spin orientation, as would occur when two black
holes coalesce (Dennett-Thorpe et al. 2002),
and their numbers are consistent with a coalescence
rate that is roughly equal to the galaxy merger rate
(Merritt \& Ekers 2002).
The almost complete lack of correlation between jet directions
in Seyfert galaxies and the angular momenta of their disks 
(Ulvestad \& Wilson 1984; Kinney et al. 2000)
also suggests that black hole coalescences were common in the past.
There are few if any ``smoking gun'' detections of 
binary black holes among AGN (e.g. Halpern \& Eracleous 2000); 
the best, but still controversial,
case is OJ 287 (Pursimo et al. 2000).

Below are discussed some additional mechanisms that have been
proposed for extracting energy from binary black holes
and hastening their decay.
When the agent interacting with the binary is stars,
continued decay of the binary implies continued destruction
of the stellar cusp.
However for most of the mechanisms discussed below, 
the detailed effects on the stellar distribution have
yet to be worked out.

\subsubsection{Scattering of stars into the loss cone}
Destruction of a pre-existing stellar cusp generates
strong gradients in the phase-space density at
$L\approx L_{lc}$, the angular momentum of an orbit
that lies just outside the loss cone.
A small perturbation can deflect a star on such an orbit
into the loss cone.
This process has been studied in detail in the context
of gravitational scattering of stars into the tidal 
disruption sphere of a single black hole 
(Lightman \& Shapiro 1977; Cohn \& Kulsrud 1978).
Once a steady-state flow of stars into the loss cone has
been set up, the distribution function near $L_{lc}$ has
the form:
\begin{equation}
f(E,L)\approx {1\over\ln(1/R_{lc})} \overline{f}(E) 
\ln\left({R\over R_{lc}}\right)
\label{eq:ls}
\end{equation}
(Lightman \& Shapiro 1977),
where $R\equiv L^2/L_c^2(E)$, $L_c(E)$ is the angular momentum
of a circular orbit of energy $E$, 
and $\overline{f}$ is the distribution function
far from the loss cone, assumed to be isotropic.
The mass flow into the central object is 
$m_*\int{\cal F}(E) dE$, where
\begin{equation}
{\cal F}(E) dE = 4\pi^2 L_c^2(E) \left\{\oint {dr\over v_r} \lim_{R\rightarrow 0} {\langle(\Delta R)^2\rangle\over 2R}\right\} {\overline{f}\over \ln(1/R_{lc})} dE.
\label{eq:relax}
\end{equation}
The quantity in brackets is the orbit-averaged diffusion coefficient
in $R$.
Yu (2002) evaluated the contribution of two-body scattering to the
decay rate of a massive binary and found that it was usually too
small to overcome the stalling that occurs when the loss cone is first
emptied.

Standard loss cone theory as applied by Yu (2002) assumes a 
quasi-steady-state
distribution of stars in phase space near $L_{lc}$.
This assumption is appropriate at the center of a globular cluster,
where relaxation times are much shorter than the age of the universe,
but is less appropriate for a galactic nucleus,
where relaxation times almost always greatly exceed a Hubble time 
(e.g. Faber et al. 1997).
The distribution function $f(E,L)$ immediately following the
formation of a hard binary is approximately a step function,
\begin{equation}
f(E,L) \approx \cases {\overline{f}(E),& $L>L_{lc}$\cr 0,&$L<L_{lc}$,\cr}
\end{equation}
much steeper than the $\sim\ln L$ dependence in a collisonally
relaxed nucleus (equation \ref{eq:ls}).
Since the transport rate in phase space is proportional
to the gradient of $f$ with respect to $L$, steep gradients imply
an enhanced flux into the loss cone.
The total mass in stars consumed by the binary can exceed
the predictions of the standard model by factors of a few,
implying greater cusp destruction and more rapid decay of the binary
(Milosavljevic \& Merritt 2003b).
This time-dependent
loss cone refilling might be particularly effective in a nucleus
that that continues to experience mergers or accretion events, in such
a way that the loss cone repeatedly returns to an unrelaxed state
with its associated steep gradients.

\subsubsection{Re-ejection}

Unlike the case of tidal disruption of stars by a single black hole, 
a star that interacts with a massive binary remains inside the galaxy and 
is available for further interactions.
In principle, a single star can interact many times with the binary
before being ejected from the galaxy or falling outside the loss
cone; each interaction takes additional energy from the binary
and hastens its decay.
Consider a simple model in which a group of $N$ stars in a spherical
galaxy interact with the binary and receive a mean energy increment 
of $\langle\Delta E\rangle$.
Let the original energy of the stars be $E_0$.
Averaged over a single orbital period $P(E)$, the binary hardens
at a rate
\begin{equation}
{d\over dt}\left({GM_1M_2\over 2a}\right) = m_* {N\langle\Delta E\rangle\over P(E)}.
\end{equation}
In subsequent interactions, the number of stars that remain inside
the loss cone scales as $L_{lc}^2\propto a$ while the ejection energy
scales as $\sim a^{-1}$.
Hence $N\langle\Delta E\rangle\propto a^1a^{-1}\propto a^0$.
Assuming the singular isothermal sphere potential for the galaxy,
one finds
\begin{equation}
{a_h\over a(t)} \approx 1 + {\mu\over M_{12}}\ln\left[ 1 + {m_*N\langle\Delta E\rangle\over 2\mu\sigma^2}{t-t_h\over P(E_0)}\right]
\end{equation}
(Milosavljevic \& Merritt 2003b).
Hence the binary's binding energy increases as the logarithm of the time,
even after all the stars in the loss cone have interacted at least 
once with the binary.
Re-ejection would occur differently in nonspherical galaxies
where angular momentum is not conserved and ejected stars could miss
the binary on their second passage. However there will generally exist
a subset of orbits defined by a maximum pericenter distance $\lap a$
and stars scattered onto such orbits can continue to interact with 
the binary.

\subsubsection{Chaotic loss cones}
Loss cone dynamics are qualitatively different in 
non-axisymmetric (triaxial or bar-like) potentials,
since a much greater number of stars may be on 
``centrophilic'' orbits which take them near to the
black hole(s) (Gerhard \& Binney 1985).
Triaxial models need to be taken seriously given recent
demonstrations (Poon \& Merritt 2002, 2003) that
black hole nuclei can be stably triaxial and that
the fraction of mass on centrophilic -- typically
chaotic -- orbits can be large.
The frequency of pericenter passages, $r_{\rm peri}<d$,
for a chaotic orbit of energy $E$ in a triaxial black-hole nuclus
is roughly linear in $d$, $N(r_{\rm peri}<d)\approx d\times A(E)$
(Merritt \& Poon 2003).
The total rate at which stars pass within a distance ${\cal R}a$ of
the massive binary is therefore
\begin{equation}
{dM\over dt} \approx {\cal R}a\int A(E)M_c(E)dE
\end{equation}
where $M_c(E)$ is the mass on centrophilic orbits.
The implied feeding rate can be comparable to that
in a spherical nucleus with a constantly-refilled loss cone,
and orders of magnitude greater than in a
loss cone that is re-supplied via star-star interactions
(Merritt \& Poon 2003).
Even transient departures from axisymmetry, for instance
during mergers, might result in substantial loss cone 
refilling due to this mechanism.

\subsubsection{Multiple black holes}

If binary decay stalls, 
an uncoalesced binary may be present 
in a nucleus when a third black hole, or a second binary,
is deposited there following a subsequent merger.
The multiple black hole system that forms will engage in its 
own gravitational slingshot interactions, 
eventually ejecting one or more of the
black holes from the nucleus (though probably not from the galaxy).
This process has been extensively modelled assuming a fixed
potential for the galaxy (e.g. Valtaoja, Valtonen \& Byrd 1989;
Mikkola \& Valtonen 1990; Valtonen et al. 1994).
The effect on the stellar distribution of $N\gap 2$, interacting 
black holes is not well understood, though $N$-body simulations
with $10-20$ massive particles and a ``live'' background 
show that the
black holes displace $\sim 10$ times their own mass before
being ejected (Merritt \& Milosavljevic 2003).
The separation and eccentricity of the dominant binary
can change dramatically during each interaction
and this may be an effective way to shorten the gravity wave 
coalescence time.
In a wide, hierarchical triple, 
the eccentricity of the dominant binary
oscillates through a maximum value of $\sim \sqrt{1-5\cos^2i/3}$,
$|\cos i|<\sqrt{3/5}$,
with $i$ the mutual inclination angle (Kozai 1962).
Blaes, Lee \& Socrates (2002) estimate that the coalescence
times in equal-mass, hierarchical triples can be reduced by factors of
$\sim 10$ via the Kozai mechanism.

\subsubsection{Gas dynamics}

If the inner $\sim 1$ pc of the nucleus contains a mass
in gas comparable to $M_2$,
torques from the gas will cause the orbit of the smaller black hole
to decay in a time of order the gas accretion
time (Syer \& Clarke 1995; Ivanov, Papaloizou \& Polnarev 1999).
Given standard assumptions about accretion disk viscosities,
the gas-dynamical decay rate would exceed that from 
gravity wave emission for $a>a_{acc}$, where
\begin{equation}
{a_{acc}\over a_h} \approx 1\times 10^{-3} \left({p\over 0.1}\right)^{2/5} \left({\sigma\over 200\ {{\rm km}\ s}^{-1}}\right)^2
\end{equation}
(Armitage \& Natarajan 2002).
If the orbit of the secondary is strongly inclined with respect 
to the accretion disk around the larger black hole,
its passages through the disk could generate periodic outbursts, 
and this has been suggested as a model
for the $\sim 12$ yr cycle of optical flaring observed
in the blazar OJ 287 (Lehto \& Valtonen 1996).
Gas deposition of the required magnitude almost certainly
occurred during the quasar epoch, although it is less
clear that this mechanism is effective for galaxies in 
the current universe.

\section{$N$-Body Studies}

\begin{figure*}[t]
\includegraphics[width=7.in,angle=0.]{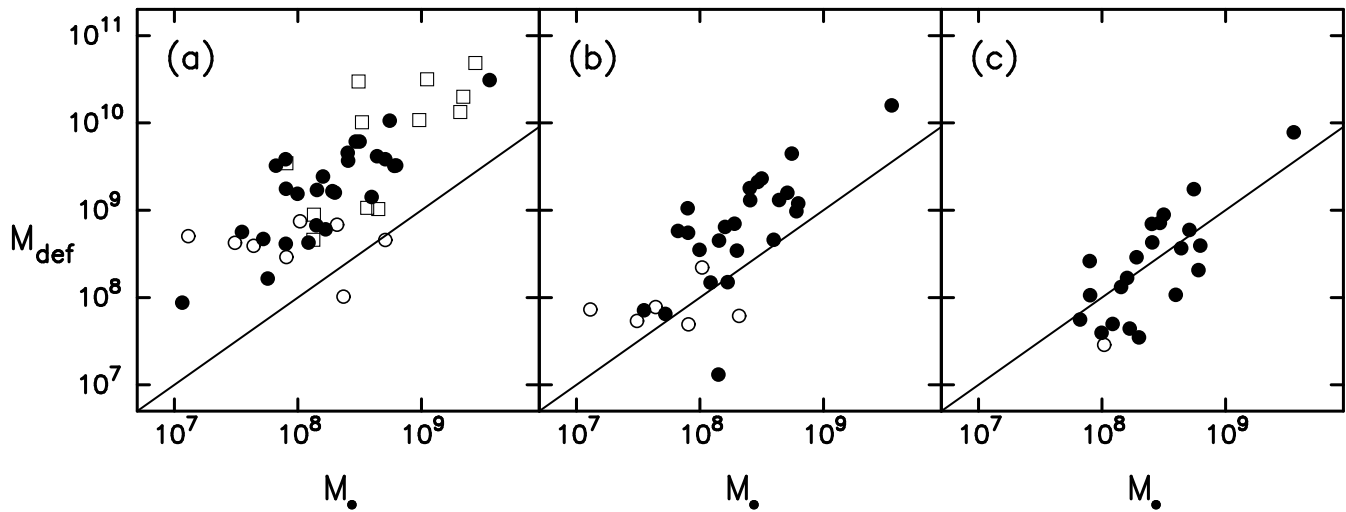}
\epsscale{1.0}
\caption{
Mass deficit versus black hole mass for three different
assumed values of $\gamma_0$, the logarithmic slope of the density
cusp before energy input from the black holes.
(a) $\gamma_0=2$; (b) $\gamma_0=1.75$; (c) $\gamma_0=1.5$.
Units are solar masses.
Solid lines are $M_{def}=M_\bullet$ (Milosavljevic et al. 2002).
}
\label{fig:four}
\end{figure*}

Unless great care is taken, $N$-body studies of binary black 
hole dynamics are unlikely to give an accurate picture of the
evolution expected in real galaxies.
This follows from the result (\S~\ref{sec:bbh_ingals}) that time scales for 
two-body scattering of stars into the binary's loss cone
are of order the Hubble time or somewhat longer in real galaxies.
In $N$-body simulations, relaxation times are shorter 
by factors of $\sim N/10^{11}$ than in real galaxies, 
hence the long-term evolution
of the binary is likely to be dominated by spurious
loss cone refilling,
Brownian motion of the black holes
and other noise-driven effects
(Milosavljevic \& Merritt 2003b).
The stalling that is predicted in the absence of
loss cone refilling in real galaxies (Figure~\ref{fig:four})
can only be reproduced via $N$-body codes
if the mean field is artificially smoothed
and the black hole binary is ``nailed down''
(e.g. Quinlan \& Hernquist 1997).

$N$-body studies are most useful at characterizing the
early stages of binary formation and decay, or 
simulating the disruptive effects of a single black hole on an
infalling galaxy; indeed scattering experiments (\S~\ref{sec:dmbs}) 
are almost useless in these regimes.
Due to algorithmic limitations, most $N$-body studies 
(e.g. Ebisuzaki, Makino \& Okumura 1991; Makino et al. 1993;
Governato, Colpi \& Maraschi 1994; Makino \& Ebisuzaki 1996; 
Makino 1997; Nakano \& Makino 1999a,b; Hemsendorf, Sigurdsson \&
Spurzem 2002)
have been based on galaxy models with unrealistically
large cores.
$N$-body merger simulations using
realistically dense initial conditions 
(Holley-Bockelmann \& Richstone 2000; 
Merritt \& Cruz 2001; Merritt et al. 2002) show that
the black hole in the larger galaxy is efficient at tidally disrupting
the steep cusp in the infalling galaxy, producing a remnant
with only slightly higher central density than that of the
giant galaxy initially.
This result helps explain the absence of dense cusps in
bright galaxies (Forbes, Franx \& Illingworth 1995)
and the persistence of the ``core fundamental plane''
in the face of mergers (Holley-Bockelmann \& Richstone 1999).

Quinlan \& Hernquist (1997) studied the evolution of a
black hole binary inside cuspy models with $\rho\sim r^{-1}$ 
and $\rho\sim r^{-2}$ and a range of black hole masses and
particle numbers, $N\le 2\times 10^5$.
Their $N$-body code was unable to simulate an actual merger
and all of their detailed results were derived
from initial conditions consisting of a single galaxy into
which two ``naked'' black holes were symmetrically dropped.
This artificial starting configuration produced substantial
evolution of the cusp before the formation of the binary.
The late evolution of the binary was found to
be strongly dependent on $N$, due in part to spurious
Brownian motion of the black hole particles.
The cores that formed were
characterized by strong velocity anisotropies, 
$\sigma_t\gg\sigma_r$, due to the ejection of stars
on eccentric orbits.

Milosavljevic \& Merritt (2001) followed the evolution
of cuspy ($\rho\sim r^{-2}$) galaxy models containing
black holes, starting from pre-merger initial conditions
and continuing until the binary separation had decayed 
a factor of $\sim 10$ below $a_h$.
The initially steep nuclear cusps were converted to
shallower, $\rho\sim r^{-1}$ profiles shortly after the
black holes had formed a hard binary; thereafter
the nuclear profile evolved slowly toward shallower slopes.
The decay rate of the binary was found not to 
be strongly dependent on $N$, probably due to the
fact that the loss cone was continuously refilled
by two-body scattering in their collisional simulation
(Milosavljevic \& Merritt 2003b).
The velocity anisotropies created during formation
of the core were much milder than in the simulations
of Quinlan \& Hernquist (1997), similar in magnitude to
the anisotropies predicted by the adiabatic growth model
(Figure~\ref{fig:one}).

An $N$-body test of the time-dependent loss cone dynamics
described in \S 4.2 would be possible using a collisional
code with $N\gap 10^7$
(Milosavljevic \& Merritt 2003b).
Such large particle numbers demand a novel
$N$-body algorithm coupled with special-purpose
hardware like the GRAPE-6.

\section{Observational Evidence for the Binary Black Hole Model}

Figure~\ref{fig:three} shows that a massive binary
must eject of order its own mass in reaching a separation
of $a_{gr}$ if $M_2\approx M_1$, or several times $M_2$ if
$M_2\ll M_1$.
These numbers should be interpreted with caution since:
(1) binaries might not decay as far as $a_{gr}$, or
the final stages of decay might be driven by some process
other than energy exchange with stars; (2) the definition of
``ejection'' used in Figure~\ref{fig:three} is escape of a star
from an isolated binary, and does not take into account the
confining effect of the nuclear potential; 
(3) the effect of repeated mergers on nuclear density profiles, 
particularly mergers involving very unequal-mass binaries, 
is poorly understood.
Nevertheless, a clear prediction of the binary black hole
model is that galactic mergers should result in the removal
of a mass of order $M_{12}$ from the nucleus.
As in the adiabatic growth model, we are handicapped in
testing the theory by lack of knowledge of the primordial
nuclear profiles.
A reasonable guess is that all galaxies originally had steep,
$\rho\sim r^{-2}$ density cusps, since these are generic in the 
low-luminosity ellipticals which are least
likely to have been influenced by mergers, and since
the adiabatic growth model predicts $\rho\sim r^{-2}$
(Figure~\ref{fig:one}).

The ``mass deficit'' is defined as the difference
in integrated mass between the observed density profile, and 
a $\rho\propto r^{-\gamma_0}$ profile extrapolated inward
from the break radius $r_b$:
\begin{equation}
M_{def} \equiv 4\pi\int_0^{r_b}\left[\rho(r_b)\left({r\over r_b}\right)^{-\gamma_0} - \rho(r)\right]r^2dr.
\end{equation}
Milosavljevic et al. (2002) and
Ravindranath, Ho \& Filippenko (2002) computed
$M_{def}$ in samples of 
``core''-profile elliptical galaxies.
The former authors found a good proportionality between
$M_{def}$ and $M_\bullet$, with 
$\langle M_{def}/M_\bullet\rangle\approx 1$ 
for $\gamma_0=1.5$ and 
$\langle M_{def}/M_\bullet\rangle\approx 10$ 
for $\gamma_0=2$
(Figure~\ref{fig:four}).
These numbers are within the range predicted by the binary
black hole model, given the uncertainties associated with
the effects of multiple mergers.
Ravindranath, Ho \& Filippenko (2001) computed black hole
masses using a shallower assumed slope for the $M_\bullet-\sigma$ relation,
$M_\bullet\propto\sigma^{3.75}$, and found a steeper, nonlinear 
dependence of $M_{def}$ on $M_\bullet$.

More rigorous tests of the binary black hole model will require a
better understanding of the expected form of $\rho(r)$.
As discussed above, while the best current $N$-body simulations
suggest $\rho\sim r^{-1}$ following binary formation
(Milosavljevic \& Merritt 2001), 
the simulations are dominated by noise over the 
long term.
If the decay stalls, the predicted density profile can be very
different: a ``hole'' forms inside of $\sim 3a_{stall}$
(e.g. Figure 1 of Zier \& Biermann 2001).
Central minima may in fact have been seen in the luminosity
profiles of a few galaxies (Lauer et al. 2002).

Other processes could result in energy exchange between
binary black holes and stars.
If the binary eventually coalesces, 
the gravitational radiation carries a 
{\it linear} momentum leading to a recoil of the 
coalesced black hole at a velocity
$v_{\rm recoil}\sim 10^2-10^3$ km s$^{-1}$ 
(Bekenstein 1973; Fitchett 1983; Eardley 1983),
and possibly even higher if the black holes were 
rapidly spinning prior to coalescence
(Redmount \& Rees 1989).
Recoil velocities of this order would eject the black hole from 
the center of the 
nucleus and its subsequent infall would displace stars.
Quantitative evaluation of this effect will require more accurate 
estimates of $v_{\rm recoil}$ based on fully general-relativistic 
calculations of black hole coalescence.

Equating $M_{def}$ with the mass removed by the binary black hole
is only justified if galaxy mergers leave nuclear density profiles 
nearly unchanged
in the absence of the gravitational slingshot.
This is known to be the case in equal-mass mergers
between galaxies with power-law cusps
(Barnes 1999; Milosavljevic \& Merritt 2001),
though less is known about the effects of unequal-mass
mergers.
It is difficult to see how the break radii in galaxies with 
$r_b\gg r_h$ (e.g. NGC 3640, 4168) can be accounted for 
by the binary black hole model unless 
mergers sometimes induce changes in $\rho(r)$ well beyond the 
radius of influence of the black holes.

Another unsolved problem is the persistence of steep, $\rho\sim r^{-2}$ 
nuclear profiles in fainter galaxies;
even these galaxies should have undergone
at least one merger since their formation.
One possibility is that faint ellipticals experienced their
last major merger at a time when spheroids contained
a much larger gas fraction than at present
(e.g. Kauffmann \& Haehnelt 2000), and that
binary decay was driven by gas dynamics during this
last merger.

A major focus of future work should be to calculate
the evolution of $\rho(r)$ as predicted by the various
scenarios for binary decay discussed in this article.

\vskip 1.truein

An abridged version of this article will appear in
Carnegie Observatories Astrophysics Series, Vol. 1: 
``Coevolution of Black Holes and Galaxies,'' edited by 
L. C. Ho (Cambridge: Cambridge Univ. Press).
I thank M. Milosavljevic for his detailed reading
of the manuscript.
This work was supported by NSF grants AST 00-71099 and
AST 02-06031,
by NASA grants NAG5-6037 and NAG5-9046, and by
grant HST-AR-08759 from STScI.

\end{document}